\documentclass[a4paper]{VisionStyle}
\usepackage{epsfig}

\begin{document}

\title{Comparison of XMM-Newton EPIC, Chandra ACIS-S3, 
ASCA SIS and GIS, and ROSAT PSPC Results for G21.5-0.9, 
1E0102.2-7219, and MS1054.4-0321}

\author{S.L.\,Snowden\inst{1,}\inst{2}} 

\institute{
NASA/Goddard Space Flight Center, Code 662, Greenbelt, MD 20771, USA
\and 
Universities Space Research Association }

\maketitle 

\begin{abstract}

This paper presents a ``man on the street'' view of the current status 
of the spectral cross calibration between the {\it XMM-Newton} EPIC, 
{\it Chandra} ACIS-S3, {\it ASCA} SIS and GIS, and {\it ROSAT} PSPC 
instruments.  Using publicly released software for the extraction of
spectra and the production of spectral redistribution response matrices 
and effective areas, the spectral fits of data from three astronomical 
objects are compared.  The three sources are G21.5-0.9 (a heavily absorbed
Galactic SNR with a power law spectrum), 1E0102.2-7219 (a SNR in the SMC 
with a line-dominated spectrum), and MS1054.4-0321 (a high redshift 
cluster with a thermal spectrum).  The agreement between the measured 
fluxes of the various instruments is within the $\pm10$\% range, and
is better when just {\it XMM-Newton} and {\it Chandra} are compared.  
Fitted spectral parameters are also in relatively good agreement 
although the results are more limited.

\end{abstract}

\section{Introduction}

In all X-ray observatory missions a great deal effort goes into 
the calibration of the scientific instruments with goals of an 
absolute accuracy usually better than, or much better than 10\%, depending 
on the quantity (e.g., energy scale, relative area, total flux, etc.).
The calibrations are usually based on extensive ground calibration 
data (which are never as complete as one would like) coupled with 
extensive in-flight observations of celestial objects (which are 
always problematic as nature has not seen fit to provide ideal 
calibration sources).  In addition, there is the fact that 
instrument responses can and will vary with time (e.g., the 
increasing charge transfer inefficiency, CTI, of CCDs).  Thus 
instrument calibration is therefore 
a long-term endeavor where occasionally the final step is just 
to declare victory and move on.  As a final editorial comment, the
astronomical community owes a great debt of gratitude to those 
individuals who undertake this very difficult task.

But back to the issue at hand, one practical 
way of examining the reliability 
of calibrations is to compare the results of various observations of 
celestial objects using various instruments.  This at least provides 
an estimate of the relative errors between the different instruments.
(There is an old joke 
from the early X-ray missions that nobody has ever measured the 
spectrum of the Crab as the calibrations of some instruments were 
fudged to give the accepted results.)  While simultaneous observations
of the same source by different instruments are ideal, for spectral 
calibration comparisons independent observations of spectrally constant 
sources can be substituted.  Thus distant supernova remnants and
high redshift clusters are the targets of choice.  However, there 
are problematic issues with both types of sources, and nature has 
not provided convenient ``standard candles'' for X-ray 
astronomy.  SNRs can have 
complex line spectra and those in the Milky Way which are small 
enough in solid angle to be useful are distant and therefore 
heavily absorbed.  High redshift clusters are not particularly 
bright so the photon statistics can be quite limited.

This paper will present results from three sources which provide 
useful results but all suffer from limitations noted above.  They 
are: 1) The Galactic SNR G21.5-0.9 which is heavily absorbed but 
provides a constant power law spectrum visible from $\sim 1-10$~keV. 
2) The SMC SNR 1E0102.2-7219 which suffers relatively little 
absorption but has a soft, very complicated, and line-rich spectrum. 
3)~The high redshift cluster MS1054.4-0321 which also suffers 
little absorption, has a relatively simple thermal spectrum, but
has limited photon statistics.  Not all instruments have 
observations of all of the sources, which is another limitation
for this study.

\section{Data Reduction and Analysis}
\label{ssnowden-WA2_sec:analysis}

To provide the pedestrian's view of the current status of the 
cross calibration, only publicly released software and calibration
data files have been used for this work.  For {\it XMM-Newton} EPIC
data, SAS V5.2 \\
(http://xmm.vilspa.esa.es/user/sas\_top.html) \\ 
has been used to extract source and background spectra,
create the spectral redistribution matrices (RMFs), and create the 
ancillary region files (ARFs, effective area vectors).
For {\it Chandra} ACIS-S3 data, CIAO 2.1 \\
(http://asc.harvard.edu/ciao/) \\
was used with occasional help from the scripts of 
Keith Arnaud.  Spectra for {\it ASCA} GIS and 
SIS data as well as {\it ROSAT} PSPC data were extracted, and 
RMFs (where necessary, otherwise standard RMFs from the public
calibration data base were used, ftp://legacy.gsfc.nasa.gov/) 
and ARFs were created using the HEASoft software package  \\
(http://heasarc.gsfc.nasa.gov/docs/corp/software.html). \\ 
In all cases, {\it Xspec} was used to fit the data after grouping for 
statistical purposes using {\it grppha} ({\it Xspec} and {\it grppha} 
are also part of the HEASoft software package).  

\section{The Cross Calibration}
\label{ssnowden-WA2_sec:cc}

\subsection{G21.5$-$0.9}
\label{ssnowden-WA2_sec:g21}

\begin{figure}[ht]
  \begin{center}
    \epsfig{file=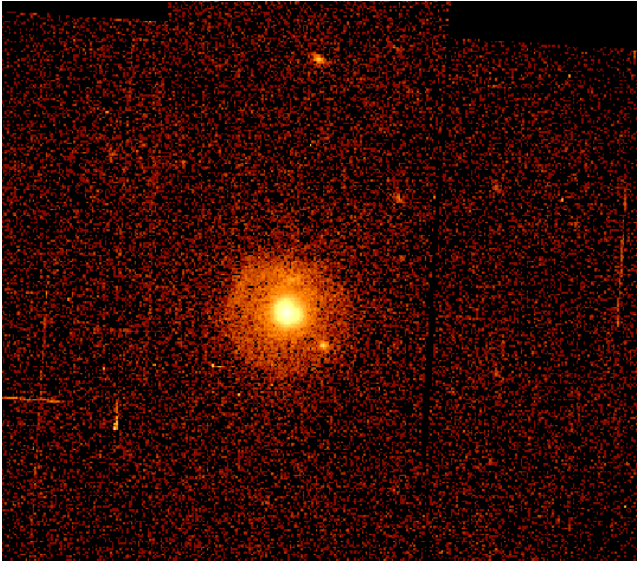, width=8.5cm}
  \end{center}
\caption{{\it XMM-Newton} EPIC MOS1 image of G21.5-0.9 from the 
Science Validation observation.}  
\label{ssnowden-WA2_fig:g21}
\end{figure}

\object{G21.5-0.9} is a Galactic SNR consisting of a Crab-like 
bright inner region and a fainter but clearly visible X-ray halo 
(Figure~\ref{ssnowden-WA2_fig:g21}).
(Note, for some of the ``science'' of this source, see the poster 
papers in these proceedings by La Palombara and Mereghetti, and 
Bocchino and Bandiera.) 
Data for this source are available from all instruments, although 
the {\it ROSAT} PSPC observation is of limited utility because the
source is so heavily absorbed.  Because of the relatively poor angular
resolution of the {\it ASCA} instruments, extraction regions large
enough to include the entire remnant were used (165\arcsec\ extraction 
radii for {\it XMM-Newton}, {\it Chandra}, and {\it ROSAT} data and 
240\arcsec\ for the {\it ASCA} data).  Source and background spectra were
extracted for all instruments.

\begin{figure}[ht]
  \begin{center}
    \epsfig{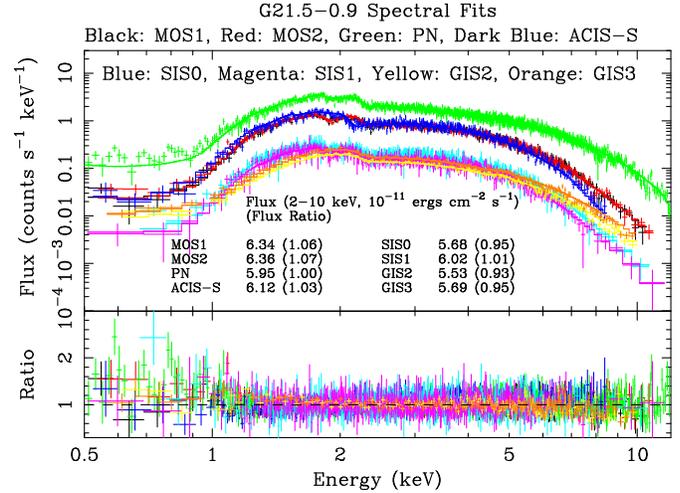}
  \end{center}
\caption{Spectral fits of the G21.5-0.9 data.  The color coding is
listed on the plot as are the fitted fluxes in the 2--10~keV band 
and the relative normalizations for the
different instruments (the PSPC results are not shown but are listed 
in Table~\ref{ssnowden-WA2_tab:summary}).}  
\label{ssnowden-WA2_fig:g21-all-spec}
\end{figure}

\begin{figure}[ht]
  \begin{center}
    \epsfig{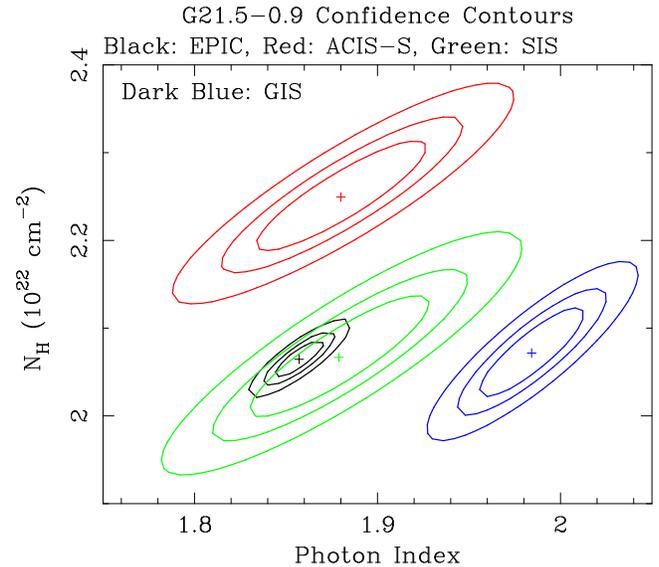}
  \end{center}
\caption{Confidence contours for the spectral parameters for fits 
to the G21.5-0.9 data.  The color coding is
listed on the plot (the PSPC results are not shown).  For this plot
the EPIC data, GIS data, and SIS data were fit together to improve 
the statistical precision.}  
\label{ssnowden-WA2_fig:g21-all-con}
\end{figure}

The data were fit over the $0.5-10.0$~keV energy
range with variation in the endpoints due to the individual 
spectral responses of the various instruments.  A simple absorbed 
power law spectrum was first fit simultaneously to the data with 
only the overall normalization being allowed to vary between the
various instruments.  The fits are displayed in 
Figure~\ref{ssnowden-WA2_fig:g21-all-spec}.  While the fits are a
bit rough below 1~keV, at higher energies they look quite good. 
(At energies below 1~keV interstellar absorption has removed most 
X-rays from the spectrum so what are typically detected are events 
which have lost some of their energy due to incomplete charge 
collection by the CCDs and electronics.)
The fitted values for the relative fluxes (scaled to the average
value) are in good agreement and range from 0.89 ({\it ROSAT} PSPC)
 to 1.07, with EPIC and ACIS-S3 values in the range 1.00 to 1.07.

\begin{figure}[!ht]
  \begin{center}
    \epsfig{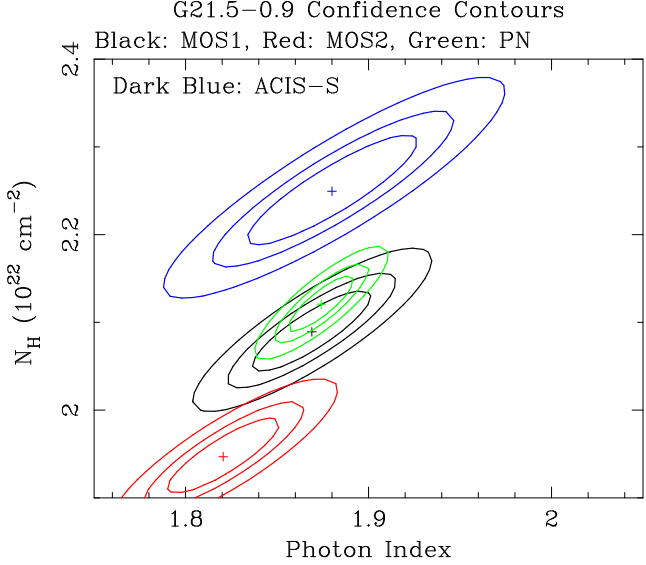}
  \end{center}
\caption{Confidence contours for the spectral parameters for fits 
to the MOS1, MOS2, PN, and ACIS-S3 G21.5-0.9 data.  The color coding 
is listed on the plot.}  
\label{ssnowden-WA2_fig:g21-epicacis-con}
\end{figure}

Figure~\ref{ssnowden-WA2_fig:g21-all-con} shows the confidence 
contours for the fitted values of the power law index and absorbing
column density.  The spectral parameters of the EPIC PN and MOS 
detectors were fit simultaneously only allowing the normalizations
to vary.  This was also done for the SIS and GIS data to improve the 
statistics.  The average results for the 
EPIC data are completely consistent
with those of the SIS.  The ACIS-S3 and EPIC slopes agree but there 
is a $\sim10$\% difference in the fitted values for the absorption 
column densities.  The GIS and EPIC results for the absorption 
column densities agree well but there is a difference of $\sim0.12$ 
in the fitted values for the slope.
Figure~\ref{ssnowden-WA2_fig:g21-epicacis-con} shows a confidence 
contour plot for the EPIC and ACIS-S3 data when the EPIC data are 
fit independently.  The PN and MOS1 values agree well while the 
MOS2 values are somewhat lower in both slope and column density.

\subsection{1E0102.2$-$7219}
\label{ssnowden-WA2_sec:e0102}

\begin{figure}[!ht]
  \begin{center}
    \epsfig{file=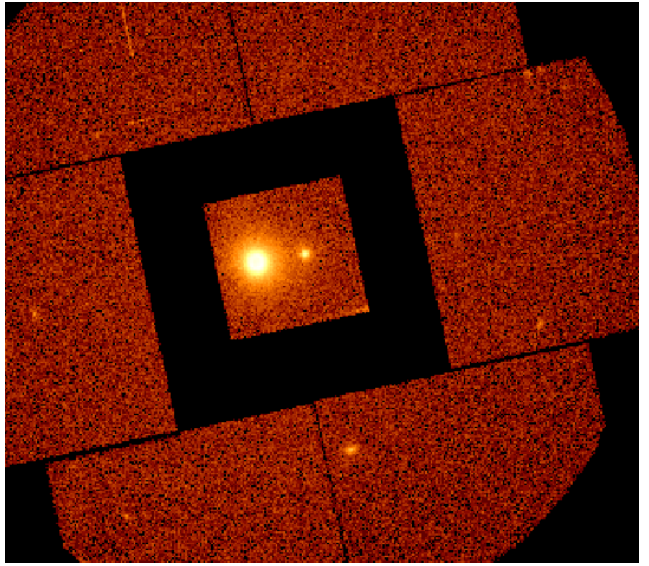, width=8.5cm}
  \end{center}
\caption{{\it XMM-Newton} EPIC MOS1 image of 1E0102.2-7219 from the 
Calibration/Performance Verification observation.}  
\label{ssnowden-WA2_fig:e0102}
\end{figure}

\object{1E0102.2-7219} is a SNR in the Small Magellanic Cloud.  It 
is beautifully resolved in the {\it Chandra} data as a shell-like 
remnant.  It's spectrum is soft and line-dominated, and very difficult 
to model short of fitting a vast number of Gaussians to the data.
Unfortunately, it was not feasible to use the PN data from the EPIC
observation as the positioning of the source for the advantage of the 
RGS caused part of the remnant to fall on a gap between the CCDs.

\begin{figure}[!ht]
  \begin{center}
    \epsfig{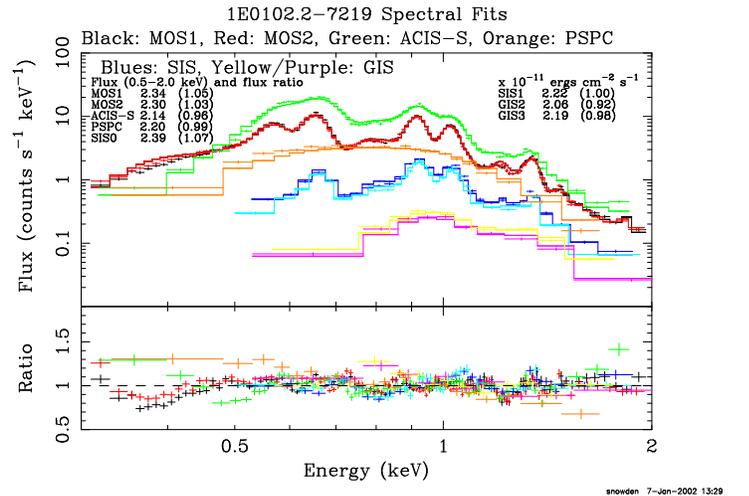}
  \end{center}
\caption{Spectral fits of the 1E0102.2-7219 data.  The color coding is
listed on the plot as are the fitted fluxes and relative normalizations 
for the different instruments.}  
\label{ssnowden-WA2_fig:e0102-all-spec}
\end{figure}

For the fits two absorbed APEC models (see 
http://hea-www.harvard.edu/APEC/) with variable abundances were
used.  The data were fit over the $0.3-2.0$~keV band, and the fit was 
not particularly significant.  However, the fits can still be used 
to compare the relative normalizations.  As can be seen in 
Figure~\ref{ssnowden-WA2_fig:e0102-all-spec} 
(and Table~\ref{ssnowden-WA2_tab:summary}), the relative 
normalizations range from 0.92 to 1.07, with the values for the
ACIS-S3 and MOS detectors ranging from 0.96 to 1.05.  

As an aside, note the 
difference between the energy resolution of ACIS-S3 (green curve in 
Figure~\ref{ssnowden-WA2_fig:e0102-all-spec}) and MOS spectra 
(the black and red curves) due to the differences in the response 
between backside and frontside illuminated CCDs.
  
\subsection{MS1054.4-0321}
\label{ssnowden-WA2_sec:ms}

\begin{figure}[!ht]
  \begin{center}
    \epsfig{file=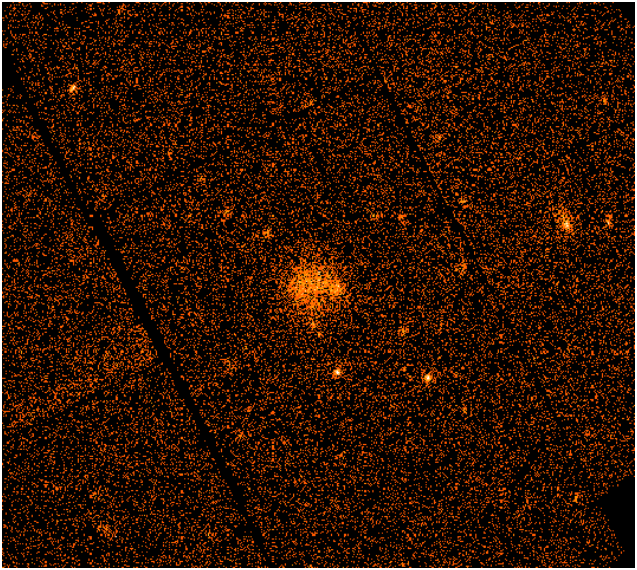, width=8.5cm}
  \end{center}
\caption{{\it XMM-Newton} EPIC MOS1 image of MS1054.4-0321 from 
the GT observation kindly provided by Mike Watson.}  
\label{ssnowden-WA2_fig:ms}
\end{figure}

\object{MS1054.4-0321} (Figure~\ref{ssnowden-WA2_fig:ms}) is a high redshift
cluster in a direction of low Galactic column density.  The limitation for 
this object as a good calibration source is its low brightness and
therefore poorer statistics.  Reasonable data are available for 
the EPIC MOS and PN, ACIS-S3, and SIS.  While the SIS data aren't 
particularly useful for constraining the spectral parameters, they do 
provide a reasonable flux comparison.
 
\begin{figure}[ht]
  \begin{center}
    \epsfig{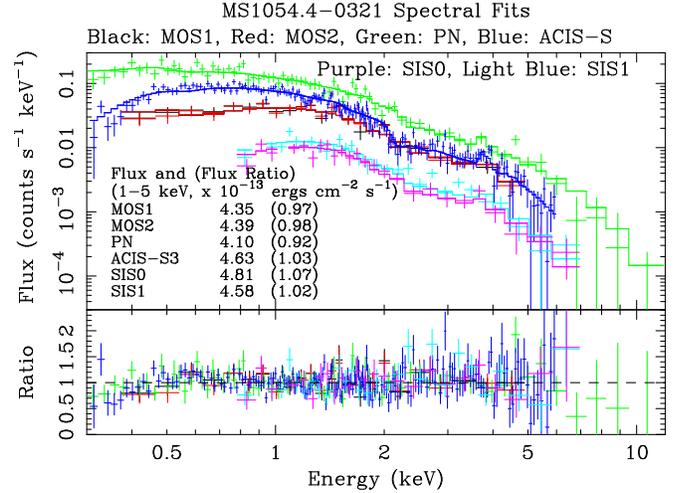}
  \end{center}
\caption{Spectral fits of the MS1054.4-0321 data.  The color coding is
listed on the plot as are the fitted fluxes and relative normalizations 
for the different instruments.}  
\label{ssnowden-WA2_fig:ms1054-all-spec}
\end{figure}

Figure~\ref{ssnowden-WA2_fig:ms1054-all-spec} shows the spectral
fits and relative fluxes for the EPIC, ACIS-S3, and SIS data.  For 
these data an absorbed thermal model (\cite{ssnowden-WA2:rs77}) was 
fit where the abundance was allowed to vary.

\begin{figure}[!ht]
  \begin{center}
    \epsfig{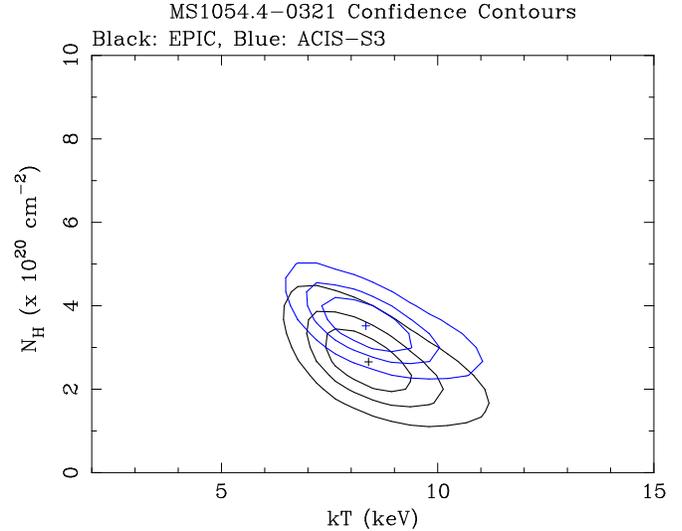}
  \end{center}
\caption{Confidence contours for the spectral parameters for the
EPIC and ACIS-S3 fits to the MS1054.4-0321 data.  The color coding is
listed on the plot.}  
\label{ssnowden-WA2_fig:ms1054-epicacis-con}
\end{figure}

For the EPIC and ACIS-S3 data, 
Figure~\ref{ssnowden-WA2_fig:ms1054-epicacis-con} shows the confidence 
contours for the fitted values for the temperature and absorption 
column density.  The EPIC data were fit simultaneously to improve the 
statistical results.  The fitted values for the parameters are 
completely consistent.

\section{CONCLUSIONS}

Table~\ref{ssnowden-WA2_tab:summary} gives a summary of the relative
flux normalizations for the simultaneous spectral fits for the three 
objects.  In all cases the full range in the {\it XMM-Newton} and 
{\it Chandra} values is better than 
$\sim10$\%, which is fairly remarkable at 
this early a stage in the missions.  When the {\it ROSAT} and 
{\it ASCA} data are included the full range is still $<20$\%.
One consistent systematic difference in the data is that the fluxes 
measured by the EPIC PN instrument are $\sim7$\% lower than the fluxes 
measured by the EPIC MOS.  This discrepancy is also seen in the 
results of Griffiths (this workshop) for the hard band, and 
both his paper and that of Haberl should be noted for their comparisons
of the EPIC MOS and PN calibrations. 

\begin{table}[ht]
\caption{Summary table of relative flux normalizations.}
\label{ssnowden-WA2_tab:summary}
\begin{center}
\footnotesize
\begin{tabular}{lccc} 
\hline \\
Object	& G21.5-0.9       & 1E0102.2-7219    & MS1054.4-0321 \\
Band	& 2.0 -- 10.0 keV & 0.5 -- 2.0 keV  & 1.0 -- 5.0 keV \\
\hline \\
MOS1    & 1.06            & 1.05            & 0.97 \\
MOS2    & 1.07            & 1.03            & 0.98 \\
PN      & 1.00            & --              & 0.92 \\
ACIS-S3 & 1.03            & 0.96            & 1.03 \\
SIS0    & 0.95            & 1.07            & 1.07 \\
SIS1    & 1.01            & 1.00            & 1.02 \\
GIS2    & 0.93            & 0.92            & --   \\
GIS3    & 0.95            & 0.98            & --   \\
PSPC    & 0.89$^{*}$      & 0.99            & --   \\
\hline  \\
\end{tabular} 
\end{center}
$^{*}$Flux compared over the 0.5 - 2.5 keV band.
\end{table} 

The cross calibration situation is also fairly good when the 
rest of the spectral parameters are considered, although the 
number of useful comparisons are much more limited.  The 
G21.5-0.9 results show that for a hard source the fitted 
values for the power law indecies are completely consistent 
to better than 0.05 ($\sim3$\%) for EPIC, ACIS-S3, and SIS data, and
agree to $\sim0.1$ when the GIS data are included.  The MS1054.4-0321 
results for EPIC and ACIS-S3 also show good agreement, but the 
statistics are much poorer.

{\bf Caveats:}  There are a number of caveats which go along with
these results.  First, the calibrations and software were current as of
the end of November, 2001.  Both the calibration and the software 
implementation for {\it Chandra} and especially for {\it XMM-Newton}
are changing with time, almost invariably for the better.  
Second, a fudge was included for the ACIS-S3 fits with a carbon 
K$\alpha$ absorption edge of optical depth 1.0 being added to
attempt to account for a recently observed systematic discrepancy in 
the area calibration.  The {\it Chandra} CIAO software and 
calibration data are being modified to include this effect.  
Third, there are clear sensitivities to
the energy range, background selection, spectral model, which
data are being fit, and what parameters are being fit 
simultaneously.  But this is expected and one of the challenges 
in trying to separate the ``calibration'' from the ``science''.
Fourth, the {\it Chandra} ACIS results are for the S3 CCD only.

As the {\it XMM-Newton} and {\it Chandra} missions progress, the 
instrument calibrations will also improve beyond the current 
levels.  With additional data, the identification of systematic 
discrepancies between the results of various instruments will 
allow the calibration teams to refine their efforts.

\begin{acknowledgements}

I would like to thank a number of people who have aided me significantly 
in the course of this work.  Dave Lumb (ESTEC) and Richard Saxton and
Steve Sembay (Leicester) with the EPIC data, Paul Plucinsky and Dick Edgar 
(CXC) and Kip Kuntz (NASA/GSFC and UMBC) with the ACIS-S3 data, and Ian
George (NASA/GSFC and UMBC) with the SIS and GIS data.

\end{acknowledgements}

\end{document}